\documentclass[sigconf]{acmart}

\AtBeginDocument{}

\setcopyright{acmlicensed} 

\acmConference[FinanSE '26]{The 3rd International Workshop on AI Solutions for Software Engineering Challenges in Financial Firms}{April 2026}{Rio De Janeiro, Brazil}
\acmBooktitle{FinanSE '26: AI for SE in Financial Firms Workshop, April 2026, Rio De Janeiro, Brazil}
\acmDOI{}
\acmISBN{}

\usepackage{multirow}
\usepackage{subcaption}
\usepackage{listings}
\usepackage{xspace}
\lstdefinelanguage{JavaScript}{
  sensitive=true,
  morekeywords={
    break,case,catch,class,const,continue,debugger,default,delete,do,else,export,extends,
    finally,for,function,if,import,in,instanceof,let,new,return,super,switch,this,throw,
    try,typeof,var,void,while,with,yield,async,await
  },
  morekeywords=[2]{Array,Boolean,Date,Error,Function,Math,Number,Object,Promise,RegExp,String,Map,Set},
  morecomment=[l]{//},
  morecomment=[s]{/*}{*/},
  morestring=[b]',
  morestring=[b]"
}

\title[Practical KPI-Based DAO Sustainability Analytics]{Operationalising DAO Sustainability KPIs: A Multi-Chain Dashboard for Governance Analytics}

\author{Silvio Meneguzzo}
\affiliation{%
  \institution{University of Turin}
  \city{Turin}
  \country{Italy}
}
\email{silvio.meneguzzo@unito.it}

\author{Claudio Schifanella}
\affiliation{%
  \institution{University of Turin}
  \city{Turin}
  \country{Italy}
}
\email{claudio.schifanella@unito.it}

\author{Valentina Gatteschi}
\affiliation{%
  \institution{Politecnico di Torino}
  \city{Turin}
  \country{Italy}
}
\email{valentina.gatteschi@polito.it}

\author{Giuseppe Destefanis}
\affiliation{%
  \institution{University College London}
  \city{London}
  \country{UK}
}
\email{g.destefanis@ucl.ac.uk}



\begin{document}

\begin{abstract}
We present DAO Portal, a production-grade analytics pipeline and interactive dashboard for assessing the sustainability of Decentralised Autonomous Organisations (DAOs) through Key Performance Indicators (KPIs) derived from on-chain governance and token events. Building on our previous work, which defined and validated a multidimensional KPI framework for DAO sustainability, this paper moves from theory to practice by operationalising that framework in software infrastructure designed for finance and FinTech contexts. The system ingests governance and treasury data from major EVM networks, harmonises the outputs, and computes sustainability scores across four dimensions: participation, accumulated funds, voting efficiency, and decentralisation. A composite 0 to 12 score is then derived using transparent thresholds that are applied client-side in the browser. 

Using a curated snapshot of more than 50 active DAOs covering 6,930 proposals and 317,317 unique voting addresses, we show how the platform surfaces recurring patterns such as persistently low participation and concentration of proposal activity. These results demonstrate how DAO Portal supports the diagnosis of governance risks and the comparison of design choices across DAOs. To promote reproducibility and adoption, we release source code, data schema, and dashboard implementation. By turning governance traces into measurable and explainable KPIs, DAO Portal provides auditable evidence of DAO sustainability and contributes software engineering infrastructure for financial applications where treasuries and decision-making rights involve significant assets.
\end{abstract}
\maketitle


\section{Introduction}
Decentralised Autonomous Organisations (DAOs) govern protocol changes and manage sizeable treasuries within DeFi and related financial ecosystems. As these organisations mature, regulators, investors, and protocol stewards require transparent, auditable, and reproducible evidence about governance health and operational sustainability. Prior studies documented participation patterns, proposer dynamics, and concentration of voting power, and highlighted measurement challenges across chains and tooling \cite{dao_survey_2024,defi_governance_empirical,governance_platforms,dao_analyzer_tool}. In our previous study, we derived a set of sustainability key performance indicators (KPIs) from theory and validated them on a curated on chain dataset \cite{meneguzzo2025kpi}. This paper moves from theory to practice by engineering DAO Portal\footnote{\emph{Live demo (test deployment)}: \url{http://daoportal.space/} (mirror: \url{http://130.192.84.45:8080/}). 
The demo serves the 50 DAO snapshot used in this paper; endpoints are read only and may be rate limited.}
, a production-oriented analytics pipeline and dashboard that computes those KPIs at scale across heterogeneous blockchains and exposes them through an explainable and reproducible interface.

Existing analytics focus on market and token activity, with limited coverage of governance specific signals such as voter participation dynamics, proposer concentration, and holder dispersion \cite{governance_platforms,dao_analyzer_tool}. Method auditability is also weak, which hampers regulator and operator use \cite{dao_survey_2024,defi_governance_empirical}. In financial contexts, where treasury control and voting outcomes condition the movement and safety of assets, operational teams face software engineering challenges that mirror those in traditional financial firms: the need for repeatable data ingestion pipelines, traceable calculations with clear provenance, and comparative views that support risk assessment, audit readiness, and regulatory review. These requirements align with established practices in financial software engineering, where reproducibility and independent verification are prerequisites for compliance.

We address this gap with a service architecture in which a backend serves harmonised governance and token metrics from pre-built JSON snapshots, and a frontend computes the sustainability KPIs. The system provides single DAO drill downs covering participation, token distribution, treasury, and proposal statistics, together with a multi DAO table offering client side sorting and banding. Using a snapshot of 50 active DAOs across major EVM networks (Ethereum, Optimism, BNB Smart Chain (BSC), Arbitrum, Polygon (Matic) Mainnet), covering 6,930 proposals and 317,317 unique voting addresses (as in \cite{meneguzzo2025kpi}), DAO Portal surfaces recurring patterns, including persistently low turnout and concentration among proposers \cite{defi_governance_empirical,dao_survey_2024}. In this way the platform acts as software engineering infrastructure for finance, turning governance traces into measurable KPI for sustainability, reproducibility, and regulatory inspection.

This paper offers four contributions:
\begin{enumerate}
    \item an operational system that implements the sustainability KPIs at scale with end to end provenance from ingestion to dashboard;
    \item a multi chain data collector for EVM governance and token events with harmonised schemas suitable for longitudinal and cross DAO analysis;
    \item a composite sustainability score (0 to 12) with transparent definitions and thresholds exposed via the API to enable consistent comparisons;
    \item reproducible artefacts, including code, schema, and example datasets, intended to support evaluation and adoption in finance and FinTech settings.
\end{enumerate}

The remainder of the paper details the system architecture (Section~\ref{sec:overview}), the data pipeline and KPI computation (Section~\ref{sec:pipeline}), the dashboard and APIs (Section~\ref{sec:ui}), the empirical evaluation (Section~\ref{sec:eval}), and a discussion of implications and limitations (Section~\ref{sec:disc} and Section~\ref{sec:lim}).

\section{Background and Related Work}
\label{sec:background}

Decentralised Autonomous Organisations (DAOs) are crypto-native organisations whose rules and assets are administered by smart contracts, with governance typically mediated by token-weighted voting. In practice, governance spans on-chain mechanisms, where proposals are encoded, voted, and executed by contracts, and off-chain signalling, where deliberation and preference aggregation occur on forums or polling systems before being bridged to on-chain execution. Common EVM patterns comprise a lifecycle of proposal creation, discussion, voting, and execution, with timelocks and access-control policies determining when and how approved actions take effect. Delegation is widely employed to concentrate voting rights in active representatives while preserving underlying ownership. Governor-style contracts exemplify on-chain voting and execution, whereas Snapshot is an instance of off-chain signalling whose outcomes may be enacted via multisignatures or timelocked controllers, depending on the DAO’s operational model \cite{dao_survey_2024, governance_platforms}.

A typical vote specifies a snapshot of voting power, a quorum or threshold criterion, a fixed voting window, and result aggregation over \textit{for}, \textit{against}, and \textit{abstain} options, before an execution step triggers one or more contract calls. Variants differ in whether abstentions count toward quorum, how proposer eligibility is determined, and whether execution is fully automated or requires human intervention. 

\subsection{Sustainability considerations for governance}
Sustainability in DAO governance involves organisational participation and legitimacy, financial capacity, and operational reliability. Persistent low turnout, proposer and delegate concentration, and underlying holder concentration create risks of plutocracy. Game-theoretic analyses of DAO voting behaviour suggest that rational self-interest can amplify these dynamics when incentive structures are misaligned \cite{11126583}. Voting-window design and quorum calibration can reinforce these dynamics. Execution pathways (autonomous timelock versus multisig mediation), upgradeability, and cross-chain fragmentation further introduce operational fragility. On-chain traces provide useful but partial proxies for these properties, since off-chain deliberation, identity ambiguity, and heterogeneous contract semantics limit comparability without normalisation and audit trails \cite{defi_governance_empirical, dao_survey_2024}.

Building on our prior work, we adopt a bounded and interpretable KPI framework for DAO sustainability, defined over four dimensions with explicit thresholds and aggregation rules \cite{meneguzzo2025kpi}.
The composite sustainability score is calculated by summing the four dimensions (Network Participation, Accumulated Funds, Voting Mechanism Efficiency, and Decentralization) with equal weight to yield a value of 0–12. Thresholds and measurement policies are defined in prior work. This paper operationalises those definitions in production software via DAO Portal, computing the constituent KPIs and the composite across multiple chains.

\subsection{Related work}
(i) Blockchain and DeFi analytics platforms provide market- and protocol-oriented dashboards that are useful for exploratory analysis but offer limited governance-specific observability. General-purpose tools such as DeepDAO aggregate treasury valuations and membership counts across DAOs, while Boardroom and Tally focus on vote aggregation and delegation tracking for on-chain governance. These platforms support discovery and participation but do not expose auditable KPI calculations, lack cross-chain schema harmonisation, and do not provide the threshold-based scoring required for reproducible sustainability assessment~\cite{governance_platforms, dao_analyzer_tool}. DAO-Analyzer offers visualisation of participation and temporal evolution but does not compute composite sustainability indicators or support multi-chain comparisons with explicit provenance.

(ii) Empirical DAO governance studies document participation patterns, proposer dynamics, concentration of voting power, and lifecycle timings across selected DAOs and protocols. These studies improve construct understanding but often rely on single-chain scopes, manual stitching of off-chain artefacts, or ad hoc denominators, which constrains cross-DAO comparability and auditability \cite{defi_governance_empirical, dao_census, dao_case_studies}. Broader ecosystem censuses confirm oligarchic tendencies and low participation, but rarely provide the governance-aware schemas required for KPI computation at scale \cite{dao_census}.

(iii) Software engineering work on blockchain monitoring addresses ingestion, normalisation, provenance, and explainability, including archive-node pipelines and schema harmonisation. These contributions underpin auditability, yet few incorporate governance-aware state machines (for example, proposal–vote–execution linkage and cross-chain message tracking) that are required to explain KPI movements and support accountability \cite{reproducible_etl, cross_chain_monitoring}.

Taken together, the literature indicates a gap: prior work does not operationalise a theory-grounded sustainability KPI framework into a multi-chain, auditable, reproducible, and explainable system and dashboard as delivered by DAO Portal. This motivates the design choices in Section~\ref{sec:overview} and the multi-chain pipeline in Section~\ref{sec:pipeline}. By encoding governance-aware semantics and provenance, DAO Portal enables comparative, audit-ready analyses.
As DAOs increasingly manage significant treasuries and decision-making rights, this type of software engineering infrastructure contributes directly to financial applications where transparency, reproducibility, and regulatory inspection are required.

\section{System Overview}\label{sec:overview}

DAO Portal provides an auditable, reproducible, and explainable pipeline that delivers DAO sustainability KPIs and a 0–12 composite score to analysts and practitioners, operationalising the framework in~\cite{meneguzzo2025kpi} within a production-ready web application.

\subsection{Architecture}
\begin{figure*}[t]
  \centering
  \includegraphics[width=1\linewidth]{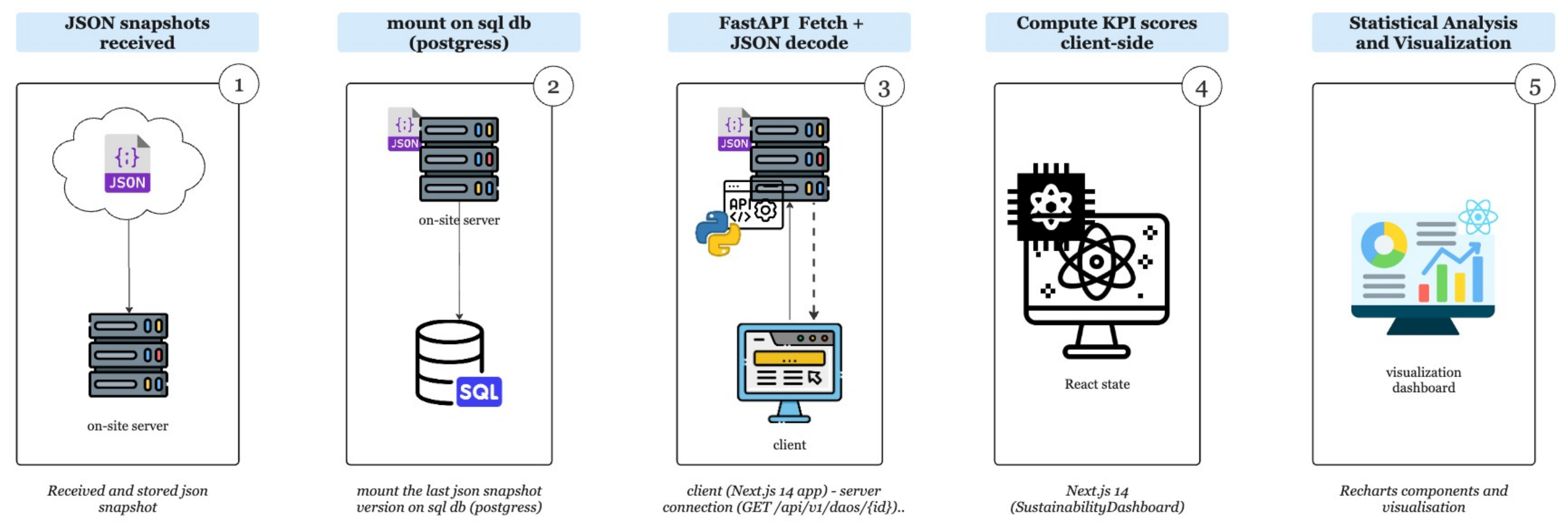}
\caption{Serving path used in this paper: harmonised JSON snapshots $\rightarrow$ FastAPI (read-only) $\rightarrow$ Next.js (client-side KPI and composite computation) $\rightarrow$ dashboard visualisation.}
  \Description{Architecture diagram showing data flow from JSON snapshots through FastAPI backend to Next.js frontend, with client-side scoring and Recharts visualisation.}
  \label{fig:arch}
\end{figure*}

DAO Portal consists of a FastAPI backend and a Next.js~14 frontend. The reference implementation is database-backed: a Celery worker can import harmonised JSON files from a mounted directory (\texttt{/data}), create a \texttt{MetricRun}, and persist one \texttt{MetricSnapshot} per metric block in Postgres. The backend exposes read-only endpoints; no server-side scoring is performed. For lightweight demonstrations, the frontend can operate in a file-backed mode by reading a bundled JSON snapshot (\texttt{web/public/dao\_data.json}) without contacting the backend. This dual mode allows operators to choose between (i) database-backed deployments with historical runs and queueable imports, and (ii) read-only snapshot mode for quick evaluation.
A schematic of the serving path is presented in Figure~\ref{fig:arch}.

\subsection{Component responsibilities}
\emph{Backend (FastAPI).} Exposes read-only endpoints that serve harmonised JSON blocks per DAO.  

\emph{Frontend (Next.js 14).} Renders single-DAO drill-downs and a multi-DAO comparison table. It applies fixed, documented thresholds to compute the four KPIs (each up to 3 points) and their 0–12 sum in the browser.  

\emph{Optional operations (private deployments).} Operators may attach a database and import jobs to persist multiple runs over time. These features are not used in the public test deployment described here.  

\subsection{Data and control flow}
\begin{enumerate}
  \item \textbf{Ingestion (optional, worker).} A Celery job reads harmonised JSON files under \texttt{/data}, records a \texttt{MetricRun}, and persists named \texttt{MetricSnapshot} blocks \\ (e.g., \texttt{network\_participation}, \texttt{accumulated\_funds}, \\ \texttt{voting\_efficiency}, \texttt{decentralisation}, \texttt{health\_metrics}).
  \item \textbf{Serving (backend).} FastAPI exposes read-only endpoints that return the latest or run-scoped blocks for one or more DAOs.
  \item \textbf{Scoring and visualisation (frontend).} The browser fetches the blocks, computes KPI scores and the 0–12 composite with fixed thresholds, and renders single- and multi-DAO views. In demo mode, the frontend consumes \\ \texttt{web/public/dao\_data.json}.
\end{enumerate}

\subsection{API surface (v1)}
\label{sec:api}
The system exposes a minimal, stable API surface used by the UI:
\begin{itemize}
  \item \texttt{GET /api/v1/daos}: paginated list of DAOs with metadata.
  \item \texttt{GET /api/v1/daos/\{id\}/enhanced\_metrics}: the five canonical metric blocks for one DAO.
  \item \texttt{GET /api/v1/daos/metrics/multi?dao\_ids=...}: metric blocks for multiple DAOs, optimised for the comparison table.
\end{itemize}
Additional endpoints support run-scoped retrieval, historical queries, and import triggering; full API documentation is provided in the source repository.

\subsection{Data model and deployment}
Each import creates a \texttt{MetricRun} (with timestamp and source file path) and one \texttt{MetricSnapshot} per metric block, supporting run-scoped retrieval and historical queries. 
The public test deployment runs FastAPI and the Next.js frontend in containers, following deployment patterns validated in other production blockchain systems~\cite{11126861}. Operators can optionally add a database for longitudinal tracking in private installations.

\section{Data Pipeline and KPI Computation}\label{sec:pipeline}

\subsection{Scope and division of labour}
DAO Portal separates \emph{extraction} from \emph{serving and analytics}. The external ETL (from our prior work~\cite{meneguzzo2025kpi}) connects to archive nodes across EVM chains, decodes governance and token events via ABIs, resolves proposal lifecycles, and harmonises outputs into JSON snapshots. DAO Portal ingests these snapshots and serves them, while the dashboard computes the sustainability KPIs and the composite score client-side and exposes them through 
widgets.

This separation keeps the web stack lightweight and auditable. The repository does not include Web3 providers or ABIs, as crawling and decoding are handled upstream. The backend focuses on deterministically transforming harmonised inputs into versioned and explainable KPI outputs.

\subsection{Snapshot contract and directory layout}
The importer accepts harmonised JSON with canonical KPI blocks either at the root or under a nested \texttt{metrics} object. In this paper, we use five blocks: \texttt{network\_participation}, \texttt{accumulated\_funds}, \\ \texttt{voting\_efficiency}, \texttt{decentralisation}, and \texttt{health\_metrics}, which are persisted as \texttt{MetricSnapshot} rows. The frontend can also consume the same structure directly from a bundled file \\ (\texttt{web/public/dao\_data.json}) in demo mode\footnote{The listing shows the primary fields used for KPI computation. Additional fields (e.g., \texttt{token\_distribution}, \texttt{health\_metrics}) and legacy keys are documented in the repository. Our evaluation uses the five canonical blocks}.

\begin{lstlisting}[basicstyle=\ttfamily\small]
{
  "dao_name": "Uniswap",
  "chain_id": 1,
  "timestamp": "2025-04-06T17:38:34.119947",
  "network_participation": {
    "num_distinct_voters": 21527,
    "total_members": 393314,
    "participation_rate": 5.4732,
    "unique_proposers": 34
  },
  "accumulated_funds": {
    "treasury_value_usd": 2.087864e9,
    "circulating_supply": 6.28494e8,
    "total_supply": 1.0e9,
    "circulating_token_percentage": 62.8494
  },
  "voting_efficiency": {
    "total_proposals": 83,
    "approved_proposals": 57,
    "approval_rate": 68.67,
    "avg_voting_duration_days": 6.07
  },
  "decentralisation": {
    "largest_holder_percent": 37.15,
    "on_chain_automation": "Yes",
    "proposer_concentration": 40.96
  }
}
\end{lstlisting}
\vspace{-0.3cm}
\paragraph{Mapping to API payload}
The backend serves harmonised JSON snapshots directly to the UI. The browser renders tiles and charts from the five canonical blocks and computes per-KPI scores locally. Table~\ref{tab:mapping} summarises the mapping between snapshot fields, the API payload consumed by the UI, and the corresponding UI components.

\begin{table*}[t]
\centering
\caption{Snapshot $\rightarrow$ API payload (as served by \texttt{/api/v1/daos/\{id\}/enhanced\_metrics} and \texttt{/api/v1/daos/metrics/multi}) and UI consumers.}
\Description{Table mapping snapshot JSON fields to API payload paths and their corresponding UI components including KPI cards, charts, and tiles.}
\label{tab:mapping}
\begin{tabular}{@{}lll@{}}
\toprule
Snapshot field & API JSON path & Used by UI \\
\midrule
\texttt{dao\_name} & \texttt{dao\_name} (top-level) & Titles, table rows \\
\texttt{chain\_id} & \texttt{chain\_id} (top-level) & Table filters/labels \\
\texttt{timestamp} & \texttt{timestamp} (top-level) & Display/ordering (optional) \\
\texttt{network\_participation} block & \texttt{network\_participation} (top-level) & KPI card \& bar chart; scoring \\
\texttt{accumulated\_funds} block & \texttt{accumulated\_funds} (top-level) & KPI card \& pie chart; scoring \\
\texttt{voting\_efficiency} block & \texttt{voting\_efficiency} (top-level) & KPI card \& pie chart; scoring \\
\texttt{decentralisation} block & \texttt{decentralisation} (top-level) & TokenDistribution chart; scoring \\
\texttt{health\_metrics} block & \texttt{health\_metrics} (top-level) & Health tile (optional) \\
\bottomrule
\end{tabular}
\end{table*}

\subsection{Ingestion and validation (file-backed, read-only)}
The backend serves a curated set of harmonised JSON snapshots (one per DAO) produced by the upstream extractor described in our prior work. In the public test deployment, snapshots are stored under a mounted data directory and are loaded or cached by the API handlers. The service is \emph{stateless} with respect to per-run provenance; no per-run SQL tables are required.

\paragraph{Validation} On read, the service checks the presence and types of the canonical blocks: \texttt{network\_participation}, \texttt{accumulated\_funds}, \texttt{voting\_efficiency}, \texttt{decentralisation}, and optionally \\ \texttt{health\_metrics}. Missing blocks are surfaced to the UI as empty objects, and booleans or enums such as \texttt{on\_chain\_automation} are normalised to \texttt{true}/\texttt{false} or \texttt{"Yes"}/\texttt{"No"} for consistent rendering.

\paragraph{Idempotence} Snapshots are immutable within a given deployment. Replacing a snapshot file atomically updates what the endpoints serve, ensuring deterministic behaviour for the figures presented in this paper.

\subsection{Enhanced-Metrics API and Client-Side Scoring}
For interactive exploration, the UI uses two patterns: (i) a multi-DAO endpoint for comparisons and (ii) a single-DAO endpoint for detail.

\paragraph{Comparison (multi-DAO)}
The multi-DAO endpoint returns the five metric blocks for each requested DAO. The \\ \texttt{SustainabilityDashboard} consumes this payload, computes four KPI scores client-side, and renders the 0–12 composite.

\paragraph{Detail (single-DAO)}  
The single-DAO endpoint returns the five metric blocks for one DAO. Run-scoped and historical payloads are also available (see Section~\ref{sec:api}).

\paragraph{Client-side scoring}
The browser applies fixed thresholds to compute the four KPIs and their 0–12 sum. The thresholds (Table~\ref{tab:score-thresholds}) are derived from our prior empirical work~\cite{meneguzzo2025kpi}; the scoring functions are part of the open-source UI.

\subsection{Derived Indicators}\label{sec:derived-indicators}
From the JSON blocks returned by the API, the UI derives a set of indicators directly in the browser, following the logic implemented in the open-source components:

\begin{itemize}
  \item \textbf{Turnout (\%)}: $\textsf{turnout} = 100 \cdot \frac{\texttt{num\_distinct\_voters}}{\texttt{total\_members}}$ if both fields $>0$; otherwise the participation KPI defaults to the lowest score.\footnote{If the computed value exceeds $100\%$, the UI treats it as anomalous and assigns the lowest participation score (see \S\ref{sec:ui}).}
  \item \textbf{Approval (\%)}: If \texttt{approval\_rate}$>1$, it is treated as a percentage; otherwise it is multiplied by $100$.
  \item \textbf{Circulating token share (\%)}: If both \texttt{circulating\_supply} and \texttt{total\_supply} $>0$, the UI computes $100 \cdot \frac{\texttt{circulating\_supply}}{\texttt{total\_supply}}$; otherwise it falls back to \texttt{circulating\_token\_percentage}.
  \item \textbf{Relative treasury (\% of circulating market cap)}: \\ If \texttt{circulating\_supply}$>0$ and \texttt{token\_price\_usd}$>0$, the UI computes:\\ $100 \cdot \frac{\texttt{treasury\_value\_usd}}{\texttt{circulating\_supply}\cdot\texttt{token\_price\_usd}}$; otherwise it defaults to $0$ and relies on absolute bins.
\end{itemize}

\subsection{KPI Scoring Thresholds (UI Policy)}\label{sec:ui}
The browser maps these indicators to four KPI scores (each up to 3 points) using fixed thresholds implemented in the React code. These thresholds were derived from empirical analysis in our prior work~\cite{meneguzzo2025kpi}, where we examined the distribution of governance metrics across active DAOs to identify natural breakpoints. The current bins are calibrated for general-purpose comparison; domain-specific deployments may require adjusted thresholds (see Section~\ref{sec:lim}). Baselines are non-zero to avoid degenerate cases for sparse or nascent DAOs. Table~\ref{tab:score-thresholds} summarises the rules; Listing~\ref{lst:ui-thresholds} shows the abridged implementation.

\begin{table*}[t]
\centering
\caption{Client-side KPI scoring thresholds used in \texttt{SustainabilityDashboard}.}
\Description{Table showing scoring rules for four KPIs: Network Participation (1-3 points based on turnout percentage), Accumulated Funds (0.75-3 points based on treasury value and circulating percentage), Voting Efficiency (1-3 points based on approval rate and voting duration), and Decentralisation (0.6-3 points based on largest holder percentage and automation).}
\label{tab:score-thresholds}
\begin{tabular}{@{}llp{6.8cm}@{}}
\toprule
\textbf{KPI} & \textbf{Points} & \textbf{Rule (key indicators)} \\
\midrule
Network Participation & 3 & $\textsf{turnout} > 40\%$ \\
 & 2 & $10\% \le \textsf{turnout} \le 40\%$ \\
 & 1 & $\textsf{turnout} < 10\%$ or invalid (e.g., $\textsf{turnout}>100\%$ or missing denominators) \\
\midrule
Accumulated Funds & 3 & $\texttt{treasury\_value\_usd} \ge \$1\text{B}$ \\
 & 2.25 & \$100M–\$1B and $\textsf{circ\_pct} > 50\%$ \\
 & 1.5 & \$100M–\$1B and $\textsf{circ\_pct} \le 50\%$ \\
 & 1.5 & (fallback) \$10M–\$1B and $\textsf{rel\_treasury} \ge 10\%$ \\
 & 1.25 & (fallback) \$1M–\$1B and $\textsf{rel\_treasury} \ge 5\%$ \\
 & 0.75 & otherwise \\
\midrule
Voting Efficiency & 3 & $\textsf{approval}>70\%$ and $3 \le \texttt{avg\_voting\_duration\_days} \le 14$ \\
 & 2 & $30\% \le \textsf{approval} \le 70\%$ and same duration window \\
 & 1 & $<30\%$ approval or duration $<3$ or $>14$ days; or $\texttt{total\_proposals}<3$ \\
\midrule
Decentralisation & 3 & $\texttt{largest\_holder\_percent} < 10\%$ \\
 & 2.4 & $10$–$33\%$ largest holder and medium/high participation and on-chain automation=\texttt{Yes} \\
 & 1.8 & $10$–$33\%$ largest holder, otherwise \\
 & 1.2 & $33$–$66\%$ largest holder \\
 & 0.6 & $>66\%$ largest holder \\
\bottomrule
\end{tabular}
\end{table*}

\begin{lstlisting}[language=JavaScript,basicstyle=\ttfamily\small,caption={Abridged client scoring (matching the UI source).},label={lst:ui-thresholds}]
const calculateNetworkParticipation = (dao) => {
  const m = dao.network_participation?.total_members ?? 0;
  const v = dao.network_participation?.num_distinct_voters ?? 0;
  if (m === 0 || v === 0) return 1;
  const rate = (v / m) * 100;
  if (rate > 100) return 1;
  if (rate > 40) return 3;
  if (rate >= 10) return 2;
  return 1;
};

const calculateAccumulatedFunds = (dao) => {
  const t = dao.accumulated_funds?.treasury_value_usd ?? 0;
  const circ = dao.accumulated_funds?.circulating_supply ?? 0;
  const total = dao.accumulated_funds?.total_supply ?? 0;
  const price = dao.accumulated_funds?.token_price_usd ?? 0;
  const circPct = total > 0 && circ > 0 ? (circ/total)*100
                 : dao.accumulated_funds?.circulating_token_percentage ?? 100;
  const rel = circ > 0 && price > 0 ? (t / (circ*price)) * 100 : 0;

  if (t >= 1e9) return 3;
  if (t >= 1e8) return circPct > 50 ? 2.25 : 1.5;
  if (t >= 1e7 && rel >= 10) return 1.5;
  if (t >= 1e6 && rel >= 5)  return 1.25;
  return 0.75;
};
\end{lstlisting}

\paragraph{Composite and sustainability levels}
Scores are summed with equal maxima to yield $C = S^{\textsf{part}} + S^{\textsf{funds}} + S^{\textsf{vote}} + S^{\textsf{decent}} \in [0,12]$. Because baselines are non-zero (Table~\ref{tab:score-thresholds}), the practical range of $C$ is $[3.35,12]$.\footnote{Baselines: $1 + 0.75 + 1 + 0.6 = 3.35$.} The UI classifies sustainability as \emph{Low} for $C<6$, \emph{Medium} for $6 \le C < 9$, and \emph{High} for $C \ge 9$.

\paragraph{Scope of scoring (client-only)}
In this release, KPI and composite scores are computed entirely in the browser from the raw snapshot blocks. The backend does not add policy tags or versioned scoring metadata. Any such extensions would be future work and would be clearly flagged in API responses when introduced.

\subsection{Edge Cases and Guards}\label{sec:guards}
\begin{itemize}
  \item \textbf{Sparse governance activity.} If \texttt{total\_proposals}<3, the voting-efficiency KPI defaults to 1 point.
  \item \textbf{Anomalous participation.} If $\textsf{turnout}>100\%$, the participation KPI defaults to 1 (the lowest score). This conservative approach ensures that data quality issues or potential manipulation (e.g., vote-buying, Sybil activity) do not inflate the sustainability assessment score. Our future design would flag such anomalies explicitly for manual review; we discuss this in Section~\ref{sec:lim}.
  \item \textbf{Automation flag.} \texttt{on\_chain\_automation} may be boolean or the strings \texttt{"Yes"}/\texttt{"No"}. The UI treats \texttt{true} and \texttt{"Yes"} as equivalent.
  \item \textbf{Funds fallback.} If \texttt{token\_price\_usd} is missing, the relative treasury is ignored and only absolute bins and circulating-share bins apply.
\end{itemize}

\subsection{From Scores to Visuals}\label{sec:scores-to-visuals}
The \texttt{SustainabilityDashboard} computes KPI scores per DAO, sums $C$, maps the result to \emph{High/Medium/Low}, and renders: (i) colour-coded KPI columns, (ii) summary tiles (counts per level and average $C$), and (iii) sorting controls (by any KPI or overall). The \texttt{MetricsDashboard} renders single-DAO tiles and charts from the same raw blocks (participation, treasury, governance outcomes, token distribution).

\section{Evaluation}\label{sec:eval}
We evaluate DAO Portal as a software artefact along three axes aligned with its intended use: (i)~\emph{correctness} of the values shown in the dashboard with respect to the ingested snapshots; (ii)~\emph{responsiveness} of the read-heavy UI and API path; and (iii)~\emph{practitioner utility} for routine governance analytics. This evaluation focuses on functional correctness and traceability rather than synthetic load testing or formal security analysis; we discuss these scope limitations in Section~\ref{sec:lim}. Our goal is to validate that the platform faithfully implements the KPI framework from our prior study~\cite{meneguzzo2025kpi} and supports the practitioner tasks it was designed for.

\subsection{Setup}
Unless stated otherwise, we use the curated snapshot of 50 active DAOs (Section~\ref{sec:pipeline}) served as read-only JSON blocks. The stack mirrors Section~\ref{sec:overview}: a FastAPI backend serving harmonised snapshots and a Next.js~14 frontend (TypeScript, shadcn/ui, Tailwind, Recharts). The comparison view loads multi-DAO payloads; the single-DAO view fetches metrics for one DAO. In demo mode, both views can fall back to a bundled snapshot file shipped with the UI.

\subsection{Correctness: snapshot$\rightarrow$UI}
We verify that the UI presents values that are deterministically derived from the harmonised blocks delivered by the API.

\paragraph{Method}
We instrumented a lightweight checker that, for a random subset of DAOs, (a)~reads the JSON snapshot used at import time, (b)~fetches the corresponding API payload for \texttt{enhanced\_metrics}, and (c)~re-executes the exact client scoring functions used by \\ \texttt{SustainabilityDashboard} to compute per-KPI points and the overall sum (code in Listing~\ref{lst:ui-thresholds}). We also verify per widget mappings used by \texttt{MetricsDashboard}:
\begin{itemize}
  \item \emph{Participation cards and charts:} \texttt{num\_distinct\_voters}, \\ \texttt{total\_members}, \texttt{participation\_rate}.
  \item \emph{Treasury card and pie:} \texttt{treasury\_value\_usd}, \\ \texttt{circulating\_supply}, \texttt{total\_supply}, \\ derived \texttt{circulating\_token\_percentage}.
  \item \emph{Governance pie:} \texttt{approved\_proposals}, \texttt{total\_proposals} (and \texttt{approval\_rate}).
  \item \emph{Token distribution:} \texttt{decentralisation.token\_distribution}, \texttt{largest\_holder\_percent}, \texttt{on\_chain\_automation}.
\end{itemize}

\paragraph{Observations}
Across DAOs in the snapshot, (i) API payloads matched the stored JSON block structure (field names and numeric types) and (ii) recomputing scores with the client functions reproduced the numbers rendered in the table (including fractional bins for \emph{Accumulated Funds} and \emph{Decentralisation}). The guard rails in the client (for example, low participation if the computed rate $>100\%$, conservative scoring when \texttt{total\_proposals}<3) produced stable and predictable behaviour on edge cases.

\subsection{Responsiveness}
\paragraph{Client-side scoring}
Sorting, filtering, and banding are executed entirely in the browser (\texttt{useMemo} over a single list response), which keeps interactions fluid and minimises server load for exploratory analysis. The API read path returns compact JSON blocks with minimal server-side shaping. When the list is paginated, server load scales with page size; the dashboard avoids N+1 patterns by computing all KPI bins client-side. This architecture prioritises auditability and client-side verifiability over server-side optimisation.

\subsection{Practitioner Utility}
To assess whether DAO Portal supports routine governance analytics, we exercised the dashboard with representative tasks identified from practitioner workflows. A formal usability study with external participants is left to future work.
Some of the dashboard's typical tasks for analysts could include:

\noindent\textbf{T1: Screen DAOs by sustainability.} The \texttt{SustainabilityDashboard} table sorts by overall score or a single KPI, with colour-coded bins and an inline ``Research Methodology'' panel that restates thresholds.

\noindent\textbf{T2: Diagnose score drivers.} Selecting a DAO, the \texttt{MetricsDashboard} reveals the raw indicators (for example, approved versus total proposals, average voting window, largest holder share). Because the client computes scores from these values, the rationale for a high, medium, or low band is visible without leaving the page.

\noindent\textbf{T3: Compare and export.} Single DAO views provide pie and bar charts (Recharts) and summary tiles; these can be exported (SVG or PNG) from the browser for reporting.\footnote{CSV or Parquet export is left to API consumers; see Future Work.}

\subsection{Evaluation Scope}
\label{sec:eval-scope}
This is a system evaluation: we do not claim new empirical findings about DAOs here, and we avoid synthetic throughput numbers that would depend on deployment choices (container resources, network conditions, database tuning). We also do not include a formal security audit, as the system is designed for read-only analytics over public blockchain data rather than for handling private credentials or executing transactions. Instead, we show that (i)~values on screen are traceable to stored blocks, (ii)~UI interactivity is achieved by shifting scoring logic to the client, and (iii)~the ingestion-and-serving split keeps the web stack simple and auditable. Section~\ref{sec:lim} discusses additional limitations.
\section{Discussion}\label{sec:disc}
DAO Portal operationalises a compact KPI framework for DAO sustainability into a web application. Two design choices are central.

\paragraph{Transparent and explainable scoring}
The dashboard renders raw governance indicators (e.g., participation rate, approval rate, largest holder share) and applies fixed, documented thresholds to compute per-KPI scores and an overall 0–12 composite. This makes the system explainable: analysts can see why a DAO falls into a given band and verify the mapping directly, without bespoke scripts. For regulators and compliance teams assessing DAOs as financial entities, such transparency is a prerequisite for accepting automated governance analytics.

\paragraph{Separation of concerns}
Crawling, ABI decoding, and cross-chain harmonisation remain upstream in a dedicated extractor, while DAO Portal focuses on ingestion, validation, storage, and presentation. This separation supports auditability and reproducibility: the server exposes exactly what was ingested, and the client computes exactly what it displays. The design mirrors established financial software engineering practice, where clear boundaries between extraction, storage, and presentation layers are essential for operational reliability.

\paragraph{Implications and trade-offs for practice}
For protocol teams and investors, the platform supports rapid triage (High, Medium, Low by composite score), targeted diagnosis of weak governance dimensions, and evidence-based discussion of design choices such as quorum thresholds or voting windows. For risk and compliance teams, storing harmonised snapshots with timestamps enables repeatable reviews and independent recomputation. These benefits come with trade-offs: equal weighting across KPIs simplifies interpretation but may not suit all contexts, client-side scoring requires careful policy documentation to avoid drift, and the focus on on-chain data limits coverage of off-chain deliberation. These trade-offs point to future research on configurable scoring policies, integration of heterogeneous governance artefacts, and validation against financial risk models.

\paragraph{Interpretation scope}
The composite score is intended for initial screening rather than definitive assessment. Aggregation can mask important trade-offs, for example when strong financial reserves coexist with weak decentralisation. Practitioners should therefore use the composite for triage and rely on per-KPI breakdowns and raw indicators for interpretation. The relationship between sustainability scores and concrete outcomes such as governance failures or treasury losses remains unvalidated and is an open research question (see Section~\ref{sec:lim}).

\paragraph{Relevance to financial software engineering}
DAO treasuries and governance outcomes involve substantial financial resources. By applying established software engineering practices including schema harmonisation, provenance tracking, deterministic computation, and verifiable client-side scoring, DAO Portal demonstrates how governance traces can be transformed into reproducible indicators suitable for risk assessment, regulatory review, and investment analysis. This positions the system within blockchain governance research while illustrating how financial applications can benefit from auditable and reproducible analytics infrastructure.

\section{Threats to Validity and Limitations}
\label{sec:lim}

\paragraph{Extractor maturity and data quality}
The external extractor is under active development. Partial coverage, ABI mismatches, and evolving normalisation rules can introduce inconsistencies in proposal counts, voting windows, or token supply figures, which propagate to KPI calculations. To limit inflated scores, the client applies conservative guards (e.g., defaulting \emph{Voting Efficiency} to 1 point when \texttt{total\_proposals}<3). \emph{Mitigation:} We plan to expose data-quality flags in the API so that known extraction gaps are explicitly surfaced in the UI.

\paragraph{On-chain scope and identity assumptions}
DAO Portal derives KPIs exclusively from on-chain evidence. Off-chain deliberation, identity, and social context are not captured, and delegation structures are only partially represented. As a result, participation and decentralisation may be under- or over-estimated, for example when low direct turnout coexists with active delegation, or when proposer concentration reflects delegate responsibility rather than oligarchy. Metrics also rely on address-level uniqueness and do not include Sybil detection, making them vulnerable to address splitting. \emph{Mitigation:} Future work may integrate delegation graphs, identity attestations, or anomaly detection to improve robustness.

\paragraph{Snapshot-based evaluation}
Results are computed on a time-bounded snapshot. Although the platform supports multiple runs, longitudinal dynamics such as token unlocks or governance reforms are not analysed here, and the UI prioritises current values for clarity.

\paragraph{Fixed scoring policy}
KPI thresholds are fixed in the client and derived from prior empirical analysis~\cite{meneguzzo2025kpi}. While interpretable, they are not universal, and alternative weights or bins may be preferable in domain-specific settings. The current API does not expose scoring policy versions. \emph{Mitigation:} Planned extensions include configurable thresholds and weighting schemes.

\paragraph{Governance heterogeneity}
DAOs differ substantially in governance design, including execution models, quorum rules, and automation mechanisms. Applying uniform KPIs across heterogeneous designs may misrepresent context-specific controls, for example when multisig-based execution scores lower on automation. \emph{Mitigation:} DAO-type classification and type-specific scoring adjustments are potential extensions.

\paragraph{Platform scope and performance}
The system is tailored to Ethereum and major EVM networks; non-EVM ecosystems would require new extractors and possibly revised KPIs. Performance characteristics depend on deployment choices, and we do not report quantitative throughput or latency benchmarks.

\paragraph{Interpretation limits}
Anomaly guards prevent UI failures but do not correct erroneous upstream data, resulting in conservative yet potentially inaccurate scores. Moreover, the relationship between sustainability scores and real governance failures or financial losses is not empirically validated. Establishing such links requires longitudinal outcome studies beyond the scope of this work.

\section{Conclusion and Future Work}
\label{sec:conclusion}

We presented DAO Portal, a platform that transforms harmonised, governance-aware on-chain data into transparent sustainability KPIs and an interpretable 0–12 composite score. The system separates extraction from serving, stores audit-ready JSON snapshots, and renders explainable single- and multi-DAO views. By applying fixed and visible thresholds, DAO Portal enables analysts and practitioners to triage DAOs and diagnose governance drivers without bespoke analytics pipelines. Future work will focus on extending chain coverage, integrating off-chain governance and lightweight identity signals, supporting configurable scoring policies, enriching longitudinal analysis, and improving robustness through anomaly detection, Sybil-resistance signals, and DAO-type classifications. Longitudinal studies are also needed to assess whether sustainability scores correlate with governance failures or treasury losses. All source code, data schemas, and the dashboard implementation are publicly available to support reproducibility and adoption.\footnote{Repository: \url{https://github.com/smeneguz/dao-portal}} More broadly, the work demonstrates how established software engineering practices such as schema harmonisation, provenance tracking, and deterministic scoring can turn DAO governance traces into reproducible indicators suitable for risk assessment, compliance review, and investment analysis in financial and FinTech contexts.

\bibliographystyle{ACM-Reference-Format}
\bibliography{refs}

\end{document}